# Evanescent Artificial Gauge Potentials for Neutral Atoms

V E Lembessis

Department of Physics and Astronomy, College of Science, King Saud University, Riyadh 11451, Saudi Arabia

**Abstract**
We show that atoms interacting with evanescent light fields, generated at the interface of a dielectric with vacuum, experience artificial gauge potentials. These potentials depend crucially on the physical parameters which characterize the evanescent fields most notably the refractive index of the dielectric material and the angle of incidence of the laser beam totally internaly reflected at the interface. Gauge fields are derived for various evanescent light fields and for both two-level and three-level systems. The use of such artificial gauge potentials for the manipulation of atoms trapped at the interfaces is pointed out and discussed.

## 1. Introduction

The advances in cooling and trapping of atomic motion, over the last three decades or so, has led to a number of fascinating applications like Bose-Einstein Condensation (BEC) of dilute atomic gases and Atom Optics [1]. It was Feynman who first suggested the use of simple and controllable quantum systems as quantum simulators of problems that cannot be modelled by a conventional computer [2]. Recent years have seen the merging of the field of cold atoms with condensed matter physics. Feynman's suggestion comes into play here. Condensed matter effects, from high-temperature superconductivity to quantum magnetism are extremely difficult to simulate on a classical computer. The difficulty becomes more apparent in systems where the electrons are strongly interacting or in cases where the system is driven far from equilibrium and then allowed to evolve. Cold atoms could act as ideal quantum simulators for such cases since their configurations are highly tunable i.e. all the interactions and parameters involved can be enginneered to suit a given model [3]-[6].

Among the physical scenarios which can be simulated include the Hubbard model and the superfluid Mott-insulator transition [7]. These effects can be modelled by cold atoms confined in optical lattices. Other examples involve the coupling of cold neutral atoms to artificial Abelian and non-Abelian magnetic and electric fields [8]. The concept of artificial gauge fields is very important in quantum simulations. Their use has been very instructive in various applications especially in high energy physics. The reason for this is that if we wish to extend quantum simulations with neutral atoms to the regime of elementary particles effects then we are going to face a major drawback: atoms are neutral with respect to the fundamental gauge fields which play a key role in modern physics so it is of a prerequisite reason to construct such gauge fields for atoms [8].

It has been shown that when a neutral atom moves in a laser field then the center-of-mas motion of the atom mimics the dynamics of a charged particle in a magnetic field with the emergence of a Lorentz-like force [9]. The orbital magnetism of a particle with charge $q$ can be considered as a consequence of the Aharonov-Bohm phase acquired by the particle when it travels along a closed contour $C$ [10]. This phase has a geometrical nature, i.e. it does not depend on the duration needed to complete the trajectory. This means that if we would like to exhibit an artificial magnetism we must find a situation where a neutral particle, for some reason, acquires a geometrical phase when it moves along a closed contour $C$. To achieve this, researchers have exploited the Berry phase effect in atom-light interactions [11]-[12]. In this case the atom-light coupling is represented by the so-called dressed states [13], which can vary on a short spatial scale (typically the wavelength of light) and the artificial gauge fields can be quite intense. If the atom, at time $t = 0$ is prepared in a dressed state $|\chi(\mathbf{r}_0)\rangle$ and moves slowly enough it follows adiabatically the local dressed state $|\chi(\mathbf{r}_t)\rangle$. When the atom completes the trajectory $C$ it returns to the dressed state $|\chi(\mathbf{r}_0)\rangle$ having acquired a phase factor which contains a geometric component. The quantum motion of the atom is formally equivalent to that of a charged particle in a static magnetic field. Such models have been studied for different beam configurations for two-level as well as for three-level atoms [9].

In this paper we consider a two-level atom interactin with an evanescent light field generated at the interface between a dielectric material and vacuum [14]. Evanescent fields are strongly localized fields with very large field gradients so they should correspond to interesting artificial gauge potentials. Moreover these fields depend crucially on the refractive index of the dielectric material and/or the angle of incidence of the laser beam giving us more control parameters with which we can monitor the properties of the artificial magnetic fields specifically their strength and/or their spatial structure [15].

The paper is organized as follows: In Section II we outline the formalism leading to the creation of the evanescent field and its interaction with the two-level atom. In Section III we apply the theory to special types of evanescent light fields generated by a Gaussian beam and a Gaussian–Laguerre one. In Section IV we consider the interaction of a three-level atom in the «lamda» configuration with the evanescent fields created by two displaced laser beams. Section V contains our comments and final conclusions.

**2. The model for the two-level atoms**
Consider first the electric field of a laser beam traveling along the z-direction in a medium of a constant refractive index $n$. The beam is characterized by an angular frequency $\omega$ and an axial wavevector $k = nk_0$, where $k_0 = \omega/c$ is the wave vector in the vacuum. The field is taken to be polarized in the $y$ direction and can be written as

$$\mathbf{E}(x,y,z,t) = \mathbf{y}\mathbb{F}(x,y,z)e^{i\Theta(\mathbf{R})}e^{-i\omega t} \quad (1)$$

where $\mathbb{F}(x,y,z)$ is an envelop function and $\Theta(\mathbf{R})$ is a position dependent phase of the field.

This situation is realizable using a glass prism of refractive index $n$. A light field as shown in Eq. (1) can be arranged to be directed towards the glass-vacuum interface (situated at $z = 0$) at an angle of incidence $\theta$, as shown in Figure 1. When $\theta$ exceeds the critical angle the beam is totally reflected and an evanscent field appears on the vacuum side of the interface.

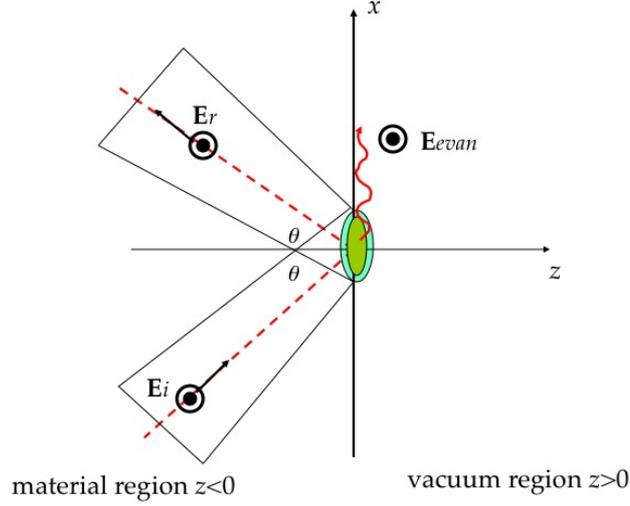

**FIG. 1**

*Creation of an evanescent light field by total internal reflection of a light beam at an angle $\theta$ (larger than the critical angle). The incident beam is arranged such thar at $\theta=0$ the beam waist coincides with the surface at z=0. The evanescent field is exponentially decaying in the direction normal to the surface.*

The electric field of this evanescent mode is given by [15],

$$\mathbf{E}_{evan}(x,y,z,t) = \mathbf{E}(x \to x\cos\theta; y; z \to -x\sin\theta, t) e^{-zk_0\sqrt{n^2\sin^2\theta-1}} e^{-ik_0 nx\sin\theta} \quad (2)$$

The atom is considered as a two-level system with a transition frequency $\omega_0$ between the ground and the excited state. It is well established [1], that the interaction of the two-level atom with the evanescent field gives rise to two dressed states, namely:

$$|\chi_1(\mathbf{R})\rangle = \begin{pmatrix} \cos(\theta(\mathbf{R})/2) \\ e^{i\phi(\mathbf{R})}\sin(\theta(\mathbf{R})/2) \end{pmatrix}, \quad |\chi_2(\mathbf{R})\rangle = \begin{pmatrix} -e^{-i\phi(\mathbf{R})}\sin(\theta(\mathbf{R})/2) \\ \cos(\theta(\mathbf{R})/2) \end{pmatrix} \quad (3)$$

with $\phi$ the position-dependent phase of the field and

$$\cos(\theta(\mathbf{R})) = \delta / \sqrt{\delta^2 + \Omega^2(\mathbf{R})}. \quad (3b)$$

Here $\delta = \omega_0 - \omega$ is the detuning of the atomic transition from the frequency of the light and $\Omega(\mathbf{R}) = |\mathbf{d} \cdot \mathbf{E}(\mathbf{R})|/\hbar$ the Rabi frequency. As has been shown, [9], under adiabaticity conditions we now have an artificial magnetic field given by

$$q\mathbf{B}(\mathbf{R}) = -\hbar\delta \frac{\Omega(\mathbf{R})}{\left(\delta^2 + \Omega^2(\mathbf{R})\right)^{3/2}} \vec{\nabla}\Omega(\mathbf{R}) \times \vec{\nabla}\varphi(\mathbf{R}) \qquad (4)$$

The artificial magnetic field corresponds to an artificial vector potential **A**. There is also an artificial scalar potential $W$ which we shall not be concerned with here although it must be taken into account when one considers the dynamics of a large sample of atoms like a BEC [9]. We will only concentrate on the magnetic fields produced by different evanescent fields created by the internal reflection of a beam at the interface between a glass prism and vacuum:

## 3. Artificial gauge potentials from evanescent light fields on two-level atoms

### A) Evanescent Gaussian fields

For a Gaussian beam the envelope function $\mathbb{F}$, appearing in Eq. 1, is given by

$$\mathbb{F}(x,y,z) = \frac{1}{\left(1 + z^2/z_R^2\right)} \exp\left(-\frac{(x^2 + y^2)}{w^2(z)}\right) \qquad (5a)$$

with $w(z)$ is the beam waist at axial position $z$. This is related to the Rayleigh range $z_R$ by:

$$w^2(z) = 2\left(z_R^2 + z^2\right)/kz_R. \qquad (5b)$$

The phase of the field, appearing in Eq. 1, is given by:

$$\Theta(\mathbf{R}) = kz \qquad (6)$$

It is easy to check, using Eq. (2), that the evanescent field corresponding to a Gaussian beam has the form,

$$\mathbf{E}_{evan}(x,y,z,t) = \mathbf{y} \frac{E_{00}}{\left(1 + x^2\sin^2\theta/z_R^2\right)} \exp\left(-\frac{\left(x^2\cos^2\theta + y^2\right)}{w_0^2\left(1 + x^2\sin^2\theta/z_R^2\right)^{1/2}}\right) e^{-zk_0\sqrt{n^2\sin^2\theta - 1}} e^{-ik_0 nx\sin\theta}$$

(7)

This field interacts with a cold two-level atom situated in the vacuum region at position $\mathbf{R} = (x, y, z)$. The resulting Rabi frequency of this interaction can be written as

$$\Omega(\mathbf{R}) = \Omega_{00} G(\mathbf{R}) \qquad (8)$$

with

$$G(\mathbf{R}) = \frac{1}{\left(1 + x^2 \sin^2\theta / z_R^2\right)} \times \exp\left(-\frac{\left(x^2 \cos^2\theta + y^2\right)}{w_0^2 \left(1 + x^2 \sin^2\theta / z_R^2\right)^{1/2}}\right) \times \exp\left(-z k_0 \sqrt{n^2 \sin^2\theta - 1}\right)$$

(9)

and $\Omega_{00}$ the Rabi frequency associated with a plane wave of the same frequency. The phase associated with this interaction is given by

$$\phi(\mathbf{R}) = x k_0 n \sin\theta. \qquad (10)$$

Inserting (9) and (10) into (4) we have for the artificial magnetic field,

$$\mathbf{B}(\mathbf{R}) = \frac{\hbar (k_0 n \sin\theta)}{q w_0} \frac{\delta \Omega^2(\mathbf{R})}{\left(\delta^2 + \Omega^2(\mathbf{R})\right)^{3/2}} \left\{ \frac{2y}{w_0 \left(1 + x^2 \sin^2\theta / z_R^2\right)^{1/2}} \hat{\mathbf{z}} - w_0 k_0 \sqrt{n^2 \sin^2\theta - 1}\, \hat{\mathbf{y}} \right\}.$$

(11)

We see that the magnetic field has two vector components, one is axial along the z-direction, i.e. perpendicular to the prism-vacuum interface, the other is along the y-direction, i.e. parallel to the interface. It can be seen from Eq. (11) that the artificial magnetic field depends crucially on important parameters namely the detuning, the Rabi frequency, the beam waist, the light wavelength and the angle of incidence.

As an illustration we explore the variations of the field components $B_y$ and $B_z$ experienced by a particle carrying a fictitious electric charge equal to the electron charge. We assume a Cesium atom, with a mass $M \approx 2.2 \times 10^{-25}\ kg$, where the evanescent field excites the transition $6^2 S_{1/2} - 6^2 P_{3/2}$ corresponding to a transition of wavelength $\lambda = 852.35\ nm$. The atom is considered as located near $z = 0$. The transition rate of the excited state is $\Gamma = 3.25 \times 10^7\ s^{-1}$. We assume a detuning of this atomic transition from the evanescent field equal to $2\Gamma$ and $\Omega_{00}$ appearing in Eq. (8) is given by $\Omega_{00} = 0.5\Gamma$. Finally the refractive index of the prism is $n=1.5$. The angle of incidence $\theta$ is $75^0$ which is greater than the critical angle.

Using the above parameters we produce Figures 2 and 3 with Figure 2 correspond to a beam waist equal to $1\,\mu m$, while Figure 3 corresponds to a beam waist equal to $10\,\mu m$.

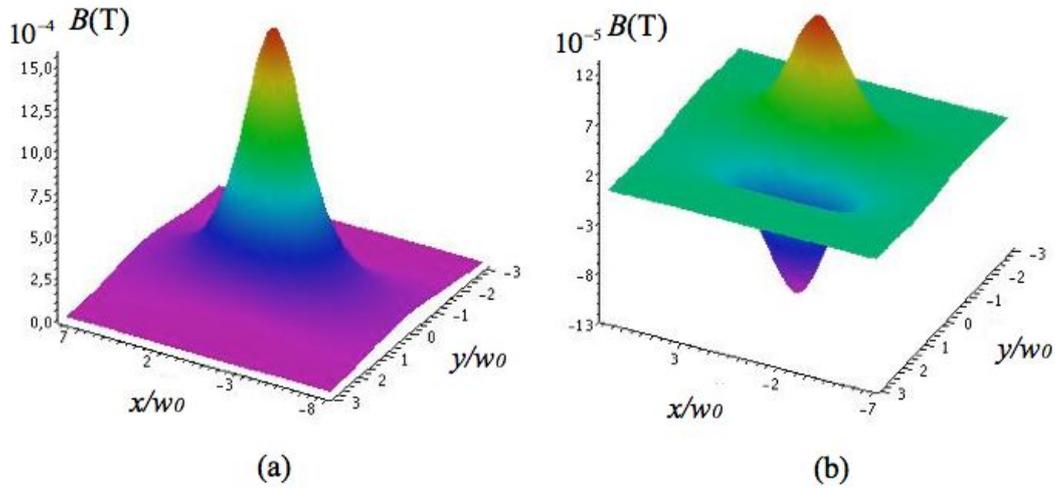

**Figure 2**
*The components of the artificial magnetic field along the y- and the z-direction (left and right respectively) in the case of a beam waist equal to $1\,\mu m$. The and y axes are scaled in $w_0$ units. The magnetic field is given in Tesla. We see that the y-component is one order of magnitude greater than the z-component. In both figures the color scale extends from magenda (minimum) to red (maximum)*

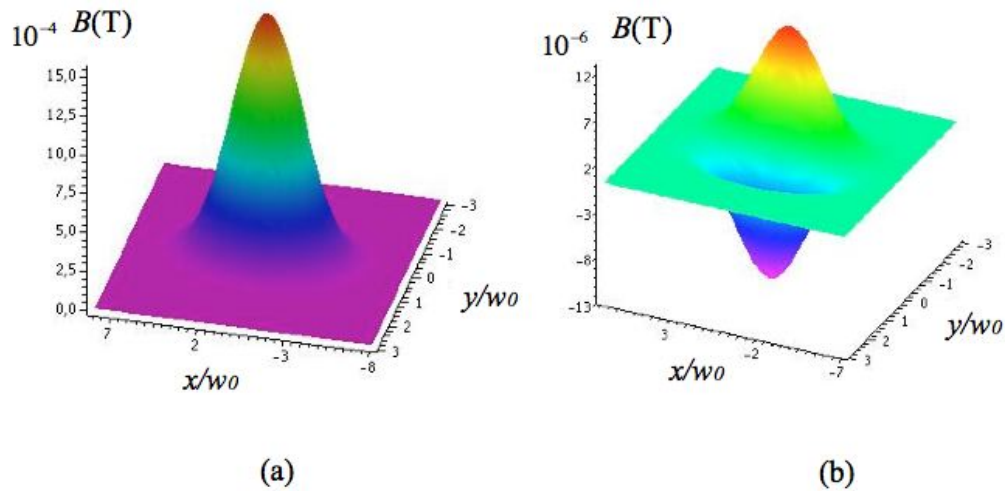

**Figure 3**
*The components of the artificial magnetic field along the y- and the z-direction (left and right respectively) in the case of a beam waist $10\,\mu m$. The and y axes are scaled in $w_0$ units. The magnetic*

*field is given in Tesla. We see that the y-component is two order of magnitude greater than the z-component. The color scale goes from magenda (minimum) to red (maximum)*

It can be seen from Figures 2 and 3 that the spatial structure of the two fields is extended in an elliptical region of space. The size and shape of this elliptical region which depends on the beam waist and the angle of incidence and follows the spatial intensity profile of the evanescent light field. Clearly the y-component dominates over the z-component. They become comparable only when the beam waist is very small which physically corresponds to very sharp field gradients.

Note that adiabaticity as a condition is important for the creation of gauge potentials [9]. Where the recoil energy is an important parameter of such a discussion. In our example here the recoil energy is $E_R = 1.4 \times 10^{-30} J$ which corresponds to an atomic recoil velocity $v_R \approx 0.36 \, cm \cdot s^{-1}$. The adiabaticity condition is satisfied provided that $\Omega(\mathbf{R}) \gg \sqrt{\Omega_{00} E_R / \hbar}$ [9]. For the above parameters this implies that $\Omega(\mathbf{R}) \gg 1.41 \times 10^{-2} \Gamma$. We have verified by explicit numerical evaluation that this condition is fully satisfied in the whole region on the x-y plane shown in Figures 2 and 3.

**B) Surface optical vortices**
It has recently been shown theoretically the possibillity that Surface Optical Vortices (SOVs) can be created at the interface between a dielectric and vacuum an optical vortex beam is internally reflected at the interface as shown schematically in Figure 1. For an optical vortex of a Gauss-Laguerre type we have for the envelope function $\mathbb{F}$ entering Eq. (1):

$$\mathbb{F}(x,y,z) = \frac{1}{(1+z^2/z_R^2)} \left[ \frac{2(x^2+y^2)}{w^2(z)} \right]^{|l|/2} L_p^{|l|} \left( \frac{2(x^2+y^2)}{w^2(z)} \right) \exp\left( -\frac{(x^2+y^2)}{w^2(z)} \right), \quad (12)$$

and the phase of the field is given by:

$$\Theta(\mathbf{R}) = kz + l\phi + \frac{k(x^2+y^2)z}{2(z^2+z_R^2)} + (2p+|l|+1)\arctan(z/z_R) \quad (13)$$

In Eq. (12) $L_p^{|l|}$ is the associated Laguerre polynomial with *l* the winding number and *p* is a node index. The last two terms in Eq. (13) are considered negligible if we work close to the beam waist or for a large Rayleigh range $z_R$. The Rabi frequency associated with the created SOV is then given by $\Omega(\mathbf{R}) = \Omega_{00} G(\mathbf{R})$ with

$$G(\mathbf{R}) = \frac{1}{\left(1 + x^2 \sin^2\theta / z_R^2\right)} \times \left[\frac{2\left(x^2 \cos^2\theta + y^2\right)}{w_0^2\left(1 + x^2 \sin^2\theta / z_R^2\right)}\right]^{|l|/2} L_p^{|l|}\left[\frac{2\left(x^2 \cos^2\theta + y^2\right)}{w_0^2\left(1 + x^2 \sin^2\theta / z_R^2\right)}\right]$$

$$\exp\left(-\frac{\left(x^2 \cos^2\theta + y^2\right)}{w_0^2\left(1 + x^2 \sin^2\theta / z_R^2\right)^{1/2}}\right) \times \exp\left(-zk_0\sqrt{n^2 \sin^2\theta - 1}\right)$$

(14)

while the phase is given by

$$\phi(\mathbf{R}) = -xk_0 n \sin\theta + l \arctan(y / x\cos\theta) \qquad (15)$$

The artificial magnetic field, in this case is given by

$$\mathbf{B}(\mathbf{R}) = B_x \hat{\mathbf{x}} + B_y \hat{\mathbf{y}} + B_z \hat{\mathbf{z}}$$

(16a)

$$B_x = \frac{\delta\Omega^2(\mathbf{R})}{q\left(\delta^2 + \Omega^2(\mathbf{R})\right)^{3/2}} k_0 \sqrt{n^2 \sin^2\theta - 1} \left(\frac{x\cos\theta}{x^2 \cos^2\theta + y^2}\right)$$

(16b)

$$B_y = \frac{\delta\Omega^2(\mathbf{R})}{q\left(\delta^2 + \Omega^2(\mathbf{R})\right)^{3/2}} k_0 \sqrt{n^2 \sin^2\theta - 1} \left(k_0 n \sin\theta + \frac{y\cos\theta}{x^2 \cos^2\theta + y^2}\right)$$

(16c)

$$B_z = \frac{\delta\Omega^2(\mathbf{R})}{q\left(\delta^2 + \Omega^2(\mathbf{R})\right)^{3/2}} \left\{\left(\frac{x\cos\theta}{x^2 \cos^2\theta + y^2}\right)\left[\frac{-x\sin^2\theta}{z_R^2\left(1 + x^2 \cos^2\theta / z_R^2\right)} + \right.\right.$$

$$+ \frac{2x}{w_0^2\left(1 + x^2 \sin^2\theta / z_R^2\right)}\left(-\cos^2\theta + \frac{\sin^2\theta\left(x^2 \cos^2\theta + y^2\right)}{z_R^2}\right) +$$

$$\frac{2x}{\left(x^2 \cos^2\theta + y^2\right)}\left(\cos^2\theta - \frac{\sin^2\theta\left(x^2 \cos^2\theta + y^2\right)}{z_R^2}\right)\left(p - (p + |l|)\frac{L_{p-1}^{|l|}(u)}{L_p^{|l|}(u)}\right)$$

$$\left.\left.\frac{|l|}{2\left(x^2 \cos^2\theta + y^2\right)}\left(2x\cos^2\theta - \frac{2x\sin^2\theta\left(x^2 \cos\theta + y^2\right)}{\left(1 + x^2 \sin^2\theta / z_R^2\right)z_R^2}\right)\right]\right\}$$

$$+\left(k_0 n\sin\theta + \frac{y\cos\theta}{x^2\cos^2\theta + y^2}\right)\left[\frac{ly}{(x^2\cos^2\theta + y^2)} - \frac{2y}{w_0^2\left(1 + x^2\sin^2\theta/z_R^2\right)}\right.$$

$$\left.+\left(\frac{2y}{x^2\cos^2\theta + y^2}\right)\left(p - (p+|l|)\frac{L_{p-1}^{|l|}(u)}{L_p^{|l|}(u)}\right)\right]\Bigg\}$$

(16d)

where the variable $u$ in the argument of the associated Laguerre polynomials is given by

$$u = \frac{2(x^2\cos^2\theta + y^2)}{w_0^2\left(1 + x^2\sin^2\theta/z_R^2\right)} \qquad (16e)$$

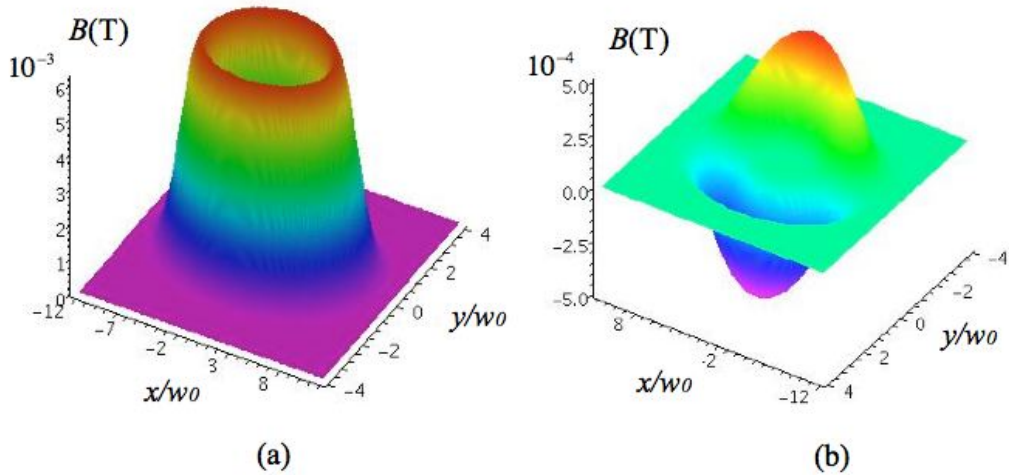

**Figure 4**
*The components of the artificial magnetic field along the y- and the z-direction (left and right respectively) for a beam with mode indices p=0, l=5 in the case of a beam waist equal to $10\ \mu m$. The and y axes are scaled in $w_0$ units. The magnetic field is given in Tesla. We see that the y-component is two order of magnitude greater than the z-component. The color scale goes from magenda (minimum) to red (maximum)*

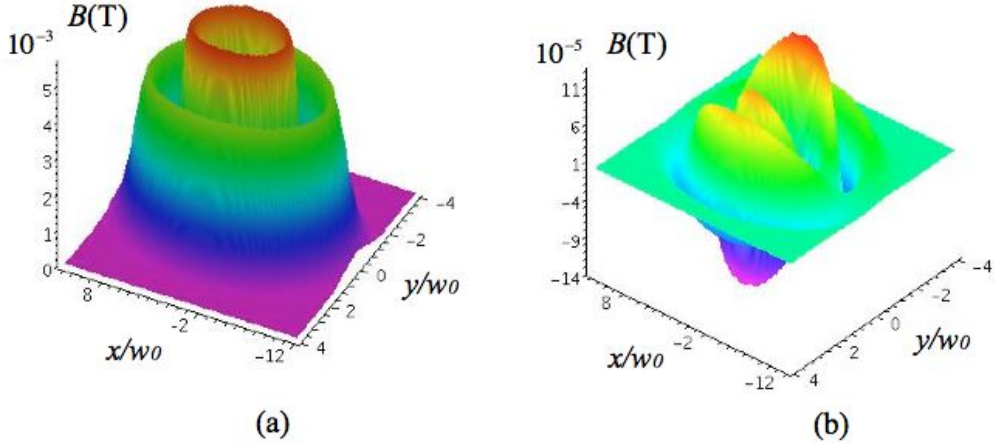

**Figure 5**
*The components of the artificial magnetic field along the y- and the z-direction (left and right respectively) for a beam with mode indices p=1, l=5 in the case of a beam waist equal to $10\ \mu m$. The*

*and y axes are scaled in $w_0$ units. The magnetic field is given in Tesla. We see that the y-component is two orders of magnitude greater than the z-component. The color scale goes from magenda (minimum) to red (maximum)*

Figures 4 and 5 show the artificial magnetic field distribution for the components along the *y* and *z* directions using the same parameters as in Figures 2 and 3. The *x*-component in both cases is not shown because is negligibly small compared with the other two. From Figures 4 and 5 we see that the magnetic field component in the *y*-direction is dominant. It also has the same spatial structure as the intensity profile of the SOV which has generated it.

As to the adiabaticity condition in the case of an optical vortex we need to remeber that processes of exchange involves both linear and orbital angular momentum. The recoil energy in this case is given by

$$E_R = \frac{(\hbar k)^2}{2M} + \frac{(\hbar l)^2}{2Mr^2} \qquad (17)$$

The second term in Eq. (17) is due to the angular momentum exchange by the atom at a radial *r* from the beam axis. It seems that it may be considerably large at small distances but in such a case we must recall that the probability of an interaction between the beam and the atom is negligibly small. The interaction probability is considerable at regions where the vortex intensity is large. For a Gauss-Laguerre beam, with $p = 0$, we know that the intensity maximizes at distances $r_{max} = w_0\sqrt{|l|/2}$ [16]. At such a distance, the ratio of the values of the two recoil kinetic energy terms is equal to $2|l|/(k^2w_0^2)$. It is easy to see that the this ratio becomes larger as either the angular momentum carried by the vortex photon becomes larger or the beam waist becomes smaller. For the parameters we have used these ratio is equal to 0.18 for a beam waist equal to 1 $\mu m$ and 0.0018 for a beam waist equal to 10 $\mu m$. So in both cases there is a small contribution from angular momentum thus our adiabaticity condition is still fulfilled.

**4. Artificial gauge potentials from evanescent light fields interacting with three-level atoms**

The use of two-level atoms for the creation of artificial gauge fields has a serious drawback arising from the fact the internal state of the atom is everywhere a linear combination of the ground and the excited state. The short lifetime of the excited state imposes a limit on the existence of such potentials. We may overcome this obstacle by considering the so called dark sates which are possible when we consider atoms with a three-level lamda configuration [17], [18]. Among the schemes which have been proposed is the case of two spatially shifted counter-propagating evanescent fields which are assumed to excite the two transitions from the ground states to the common excited state. We assume that the evanescent fields have been created by total internal reflection of two identical counter-propagating displaced Gaussian laser beams as shown in Figure 6. The beams are laterally displaced by a distance *a*.

The evanescent fields interact with the atom with the following Rabi frequencies:

$$\Omega_1(\mathbf{R}) = \frac{\Omega_{00}}{\left(1 + x^2 \sin^2\theta / z_R^2\right)} \times \exp\left(-\frac{\left(x^2 \cos^2\theta + (y-a)^2\right)}{w_0^2 \left(1 + x^2 \sin^2\theta / z_R^2\right)^{1/2}}\right) \times \exp\left(-zk_0 \sqrt{n^2 \sin^2\theta - 1}\right)$$

(18a)

$$\Omega_2(\mathbf{R}) = \frac{\Omega_{00}}{\left(1 + x^2 \sin^2\theta / z_R^2\right)} \times \exp\left(-\frac{\left(x^2 \cos^2\theta + y^2\right)}{w_0^2 \left(1 + x^2 \sin^2\theta / z_R^2\right)^{1/2}}\right) \times \exp\left(-zk_0 \sqrt{n^2 \sin^2\theta - 1}\right)$$

(18b)

The artificial magnetic field produced by such a configuration is [16],

$$\mathbf{B} = \frac{\hbar}{q} \frac{\vec{\nabla} S \times \vec{\nabla}|\zeta|^2}{\left(1 + |\zeta|^2\right)^2} , \quad (19)$$

where in our case $\zeta$ and $S$ are defined as follows:

$$\zeta = \frac{\Omega_1}{\Omega_2}, \quad S = 2xk_0 n \sin\theta. \quad (20)$$

The artificial magnetic field produced has only one "axial" component given by:

$$\mathbf{B} = -\frac{8\hbar k_0 na\sin\theta}{qw_0^2\left(1+x^2\sin^2\theta/z_R^2\right)^{1/2}}\frac{1}{\cosh^2 u}\mathbf{z} \quad (21)$$

where $u$ is defined as

$$u = \frac{a^2 - 2ay}{w_0^2\left(1+x^2\sin^2\theta/z_R^2\right)^{1/2}}. \quad (22)$$

As an illustration we consider the case of two *p*-polarized Gaussian beams internally reflected at the focus planes coinciding with the glass-vacuum interface. If the glass has a relatively high refractive index ( > 3.2) we may construct evanescent fields with almost circular polarization [17].

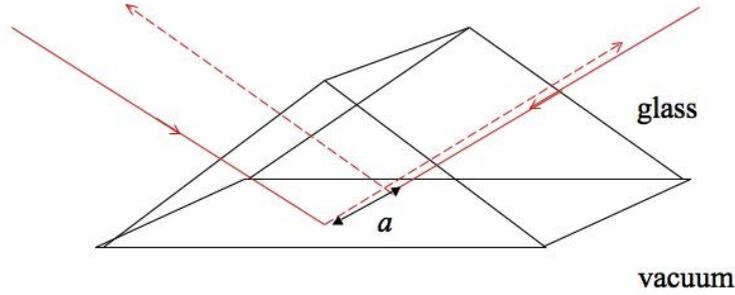

**Figure 6**

*Schematic representation of generation of two displaced counter-propagating evanescent fields.*

We assume that with these fields we could excite a Λ-scheme interaction in the D1 line of Cesium atom. The parameters are as follows: a glass index $n = 3.5$, a wavelength for the beams $\lambda = 894.6$ nm, and a beam waist of 8 $\mu m$ for both beams. The two evanescent beams are displaced in the *y*-direction by a distance $a = 0.8\, w_0$. The artificial magnetic field produced using Eq. (21) is shown Figure 7.

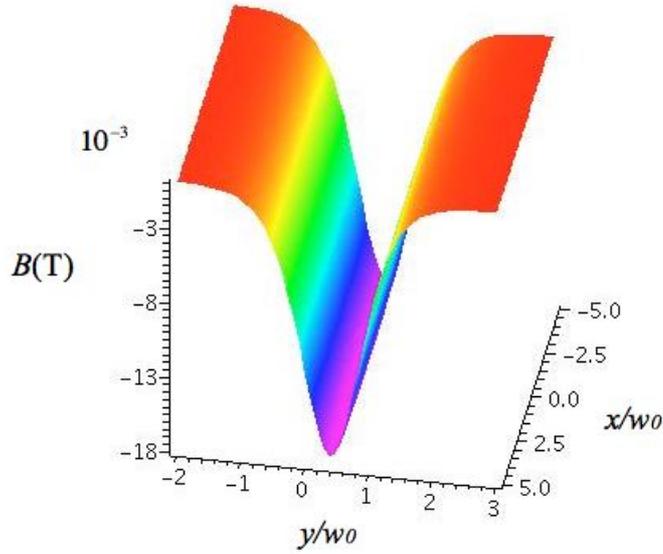

**Figure 7**
*The artificial magnetic field along the z-direction for two displaced Gaussian evanescent wave in the case of beam waists equal to . The x and y axes are scaled in units. The two beams are displaced in the y-direction by a distance . The magnetic field is given in Tesla. The color scale goes from magenda (minimum) to red (maximum)*

We see that the produced magnetic field has a considerable magnitude in the region between zero and $w_0$ (y-direction). As in the previous cases, the strength of this magnetic field is controlled by the angle of incidence, the glass refractive index and the beam waist. The magnetic field strength depends also on the parameter *a*, which the displacement between the two beams. The question of adiabaticity in the three-level atom case has been discussed in details in [18]. These authors have showed that for an atom with velocity **v** adiabaticity is satisfied provided $F \ll \Omega$, where $F = \left|\vec{\nabla}\zeta \cdot \mathbf{v}\right| / \left(1 + |\zeta|^2\right)$, and $\Omega = \sqrt{\Omega_1^2 + \Omega_2^2}$. We have again verified by explicit evaluation that condition is perfectly satisfied for atomic velocities of few centimetres per second (which is a typical speed of sound in an atomic BEC).

## 5. Comments and conclusions

Artificial magnetic fields are a new tool for the manipulation of the atomic motion and it is an area where we can see a fascinating demonstration of the idea of quantum simulation as suggested by Feynman. The traditional way for creating such fields is the rotation of the atomic system, such that the vector potential will appear in the rotating frame of reference [19]. In this case the atom feels a uniform magnetic field. However, this is an experimentally demanding technique. The use of the discrete symmetric structure of an optical lattice where we can achieve asymmetric transitions between lattice sites has been also proposed for the creation of artificial magnetic fields [20]. This method is not applicable to an atomic gas which does not constitute a lattice. In fact the light induced gauge potentials discussed here are free from all the demanding experimental techniques to investigate them and also can be used for the creation of non-Abelian gauge potentials [18].

In this work we have studied the case where evanescent fields can be exploited for the creation of such artificial surface magnetic fields. The nature of gauge fields is purely geometric, thus in nanotechnology, where the light is generated and propagates inside complex tailored structures they will be very important regarding the manipulation of atomic motion in such structures.

We have studied different cases of evanescent fields and we have shown the existence of artificial magnetic fields which, in the case of two-level atoms, have components parallel (main) and perpendicular (secondary) to the interface. The perpendicular component could be used to stimulate rotations on the x-y plane but it is weaker compared to the component along the y-direction which will allow circular in-plane trajectories. In the three-level case we have a dominant component along the z-direction which remains constant in a narrow region along the y-direction but extended along x-direction (the propagation direction of the evanescent field). This component can be used to rotate atomic particles on the surface.

In a two-level system the excited state lifetime imposes a limit on the life of such artificial magnetic fields. This is the reason researchers turned to three-level atom dark states. But in a two-level system we may have excited states with large lifetimes if we consider electrically quadrupole allowed transitions. It has been showed recently [21] that mechanical effects of laser light which interact in electric quadrupole with two-level atoms could be significant if the laser beam is in the form of a light vortex. In this case the interaction has a rich spatial structure with very sharp gradients which could result in new forms of artificial magnetic fields. However we shall not consider quadrupole allowed transitions in this context any further.

**Acknowledgments**

I would like to thank ESF for supporting a visit to the University of York through the Short Visit Grant No 5584 of the POLATOM network.

**Captions for figures**

1. Creation of an evanescent light field by total internal reflection of a light beam at an angle $\theta$ (larger than the critical angle). The incident beam is arranged such thar at $\theta=0$ the beam waist coincides with the surface at $z=0$. The evanescent field is exponentially decaying in the direction normal to the surface.
2. The components of the artificial magnetic field along the y- and the z-direction (left and right respectively) in the case of a beam waist equal to $1\,\mu m$. The and y axes are scaled in $w_0$ units. The magnetic field is given in Tesla. We see that the y-component is one order of magnitude greater than the z-component. In both figures the color scale extends from magenda (minimum) to red (maximum).

3. The components of the artificial magnetic field along the y- and the z-direction (left and right respectively) in the case of a beam waist $10\,\mu m$. The and y axes are scaled in $w_0$ units. The magnetic field is given in Tesla. We see that the y-component is two order of magnitude greater than the z-component. The color scale goes from magenda (minimum) to red (maximum).

4. The components of the artificial magnetic field along the y- and the z-direction (left and right respectively) for a beam with mode indices p=0, l=5 in the case of a beam waist equal to $10\,\mu m$. The and y axes are scaled in $w_0$ units. The magnetic field is given in Tesla. We see that the y-component is two order of magnitude greater than the z-component. The color scale goes from magenda (minimum) to red (maximum).

5. The components of the artificial magnetic field along the y- and the z-direction (left and right respectively) for a beam with mode indices p=1, l=5 in the case of a beam waist equal to $10\,\mu m$. The and y axes are scaled in $w_0$ units. The magnetic field is given in Tesla. We see that the y-component is two orders of magnitude greater than the z-component. The color scale goes from magenda (minimum) to red (maximum).

6. Schematic representation of generation of two displaced counter-propagating evanescent fields.

7. The artificial magnetic field along the z-direction for two displaced Gaussian evanescent wave in the case of beam waists equal to . The x and y axes are scaled in units. The two beams are displaced in the y-direction by a distance . The magnetic field is given in Tesla. The color scale goes from magenda (minimum) to red (maximum)